# Thermally-Switchable Metalenses Based on Quasi-Bound States in the Continuum

*Stephanie C. Malek[1], Cheng-Chia Tsai[1], and Nanfang Yu[1*]*

[1]Department of Applied Physics and Applied Mathematics, Columbia University, New York, NY 10027, United States
[*]ny2214@columbia.edu

**Abstract**

Dynamic wavefront shaping with optical metasurfaces has presented a major challenge and inspired a large number of highly elaborate solutions. Here, we experimentally demonstrate thermo-optically reconfigurable, nonlocal metasurfaces using simple device architectures and conventional CMOS-compatible dielectric materials. These metasurfaces support quasi-bound states in the continuum (q-BICs) derived from symmetry breaking and encoded with a spatially varying geometric phase, such that they shape optical wavefront exclusively on spectrally narrowband resonances. Due to the enhanced light-matter interaction enabled by the resonant q-BICs, a slight variation of the refractive index introduced by heating and cooling the entire device leads to a substantial shift of the resonant wavelength and a subsequent change to the optical wavefront associated with the resonance. We experimentally demonstrate a metalens modulator, the focusing capability of which can be thermally turned on and off, and reconfigurable metalenses, which can be thermo-optically switched to produce two distinct focal patterns. Our devices offer a pathway to realize reconfigurable, multifunctional meta-optics using established manufacturing processes and widely available dielectric materials that are conventionally not considered "active" materials due to their small thermo-optic or electro-optic coefficients.

**Introduction**

The development of tunable and reconfigurable optical metasurfaces poses a substantial challenge that has captured the interest of the optics community, but a widely favored solution has not yet emerged.[1–5] A common approach has been to apply a single stimulus to the entire device such as stretching a metasurface on a polymer substrate[6,7] or strongly modulating the refractive index of a metasurface made of active materials[8,9]. This approach has led to the demonstration of metalenses with continuously tunable focal distance[6,7], bifocal switchable metalenses[8], and reconfigurable vectorial meta-holograms[9]. Entirely reconfigurable metasurfaces that can actively mold the wavefront into any arbitrary shape require that different stimuli are applied to different portions of the device. To this end, so far there have only been experimental demonstrations of one-dimensional wavefront shaping based on meta-gratings with semiconductor quantum wells[10] and a transparent conductive oxide[11], and many of the proposals for multi-electrode tunable metasurfaces with phase-change materials[12] and lithium niobate[13,14] await experimental implementation due to the practical difficulties.

It is fundamentally difficult to drastically change the shape of the emerging wavefront by inducing a limited change to the refractive index of an entire metasurface. This is the main reason that active metasurfaces composed of conventional materials and fabricated with well-established manufacturing processes have largely proven elusive (**Table S1**). The meta-units of "local" or conventional metasurfaces are usually spectrally broadband and do not support resonant light-matter interactions. Therefore, changing the refractive index of these metasurfaces slightly cannot substantially modify the wavefront shape or device functionality. Instead, large refractive index changes as well as elaborate designs must be employed to induce even modest changes to the wavefront, which restrict the materials platform to a limited few (e.g., phase-change materials[15] and liquid crystals[9,16]). "Nonlocal" photonic structures support spatially-extended and spectrally-narrow optical modes[17,18] and can be designed to foster enhanced light-matter interactions.[19,20] These structures have been leveraged to modulate the intensity of free-space light;[21–23] however, they usually do not allow for molding the wavefront shape. A category of wavefront-shaping metasurfaces that offer more tunability than local metasurfaces are based on Huygens' metasurfaces made from highly-tunable phase change materials.[8,24,25] Huygens' metasurfaces[26,27] are resonant devices capable of shaping a wavefront through optical modes associated with Mie resonances. These modes feature lower quality factors (Q-factors), weaker light-matter interaction, and higher spatial confinement of light than nonlocal metasurfaces;[28] therefore, Huygens' metasurfaces require relatively large changes to the refractive index ($\frac{\Delta n}{n} > 20\%$) to actively tune the wavefront shape (**Table S1**).

In this work, we experimentally demonstrate nonlocal metasurfaces that, with a tuning of the refractive index by no more than a few percent, can completely modify the optical wavefront (**Fig. 1**). The physical basis of these devices are quasi-bound states in the continuum (q-BICs) that originate from period-doubling perturbations that break the in-plane symmetry of a dielectric photonic crystal (PhC) slab. The vectorial property (i.e., which incident polarization excites which mode) of these devices can be engineered through the symmetry of the unperturbed PhC lattice and the choice of perturbation, while the scalar property (i.e., Q-factor) is controlled by the strength of the perturbation.[29,30] By controlling the in-plane orientation angle of the dimerizing perturbation, selected modes with particular symmetries can introduce a geometric phase, which can be leveraged to shape the emerging wavefront.[30,31] This approach has enabled us to demonstrate theoretically[30,31] and experimentally[32] nonlocal metasurfaces that shape wavefronts only at narrowband optical resonances and act like an unpatterned substrate for non-resonant light. Nonlocal metasurfaces supporting high Q-factor q-BICs represent an appealing scheme for realizing switchable devices, which we have previously demonstrated computationally.[33] In this scheme, adjusting the refractive index of the entire metasurface shifts the resonant wavelength of the q-BIC relative to the wavelength of the narrowband incident light, and the wavefront is shaped only when the wavelength of the q-BIC and that of the source coincide. Our nonlocal metasurfaces can support multiple q-BIC modes encoded with entirely distinct wavefront shapes by incorporating several orthogonal perturbations (i.e., perturbations that break different symmetries) in the same metasurface.[31] Such multifunctional metasurfaces can enable active devices that switch between different, deliberately prescribed wavefronts when the refractive index of the entire metasurface is adjusted so that different resonant modes are successively brought into alignment with the narrowband incident light.[33] In this work, we experimentally demonstrate active nonlocal metalenses based on the thermo-optic effect of germanium in the near-infrared where lensing can be switched on and off ("metalens modulator") (**Fig. 1a**) or switched to produce distinct focal patterns (**Fig. 1b**). Furthermore, we demonstrate in simulation multiple thermo-optically switchable beam-steering metasurfaces in the visible.

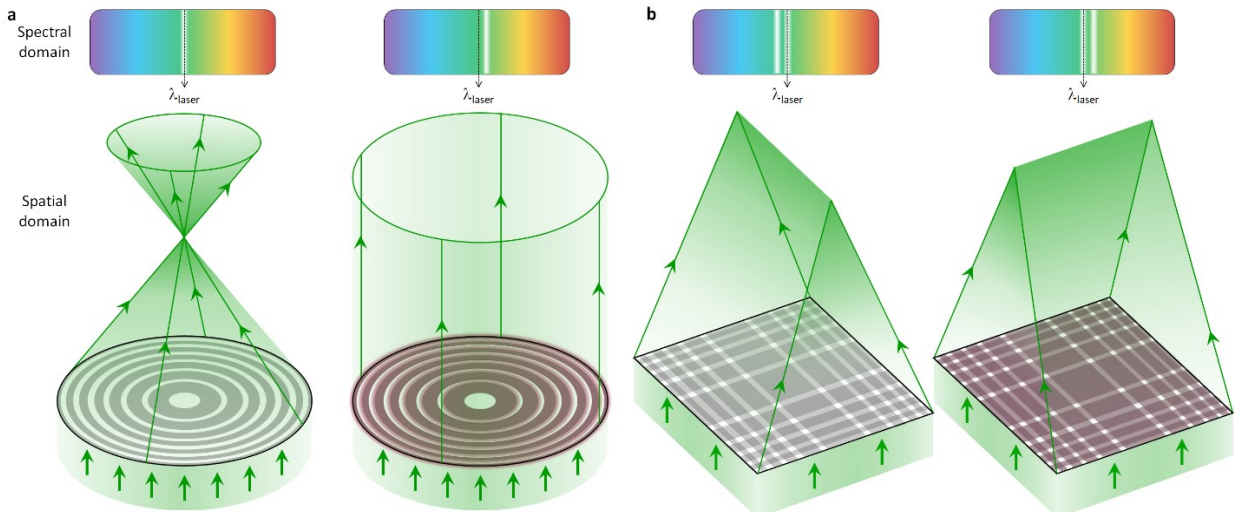

**Figure 1.** Thermally-switchable metalenses. (a) Schematic illustration of a metalens modulator, the focusing capability of which can be thermally turned on and off for spectrally narrowband excitation at $\lambda_{laser}$. (b) Schematic illustration of a reconfigurable metalens, which can be thermo-optically switched to produce two distinct focal patterns. In (a) and (b), the left and right illustrations represent, respectively, device operation at room temperature and at an elevated temperature.

**Results**

We first design and experimentally demonstrate a nonlocal metasurface modulator such that the emerging light wave can be switched between a converging spherical wave and a plane wave by changing the temperature of the entire device. A metalens modulator is unusual in the landscape of tunable metasurfaces, but has previously been realized through electrical tuning of the excitonic resonance in monolayer $WS_2$.[34] We begin by devising a nonlocal metalens in the near-infrared. Germanium is selected as the active material for its relatively high thermo-optic coefficient of $dn/dT \sim 8\times10^{-4}$ $K^{-1}$ among CMOS-compatible materials.[35] The meta-units are designed to support a q-BIC capable of imparting spatially varying geometric phase to the wavefront while also enabling substantial modal overlap with the active material. The former is achieved by choosing a meta-unit symmetry that conforms to a *p2* plane group,[31] while the latter can be controlled by the choice of the q-BIC mode (i.e., fundamental vs. higher-order mode, transverse-electric (TE) vs. transverse-magnetic (TM) mode), the meta-unit motif (i.e., apertures in a thin film vs. pillars, aperture or pillar shape), and the Q-factor of the q-BIC mode.

Our metasurface originates from an unperturbed metasurface consisting of a square lattice of square apertures defined in a germanium thin film. A period-doubling perturbation deforms each adjacent pair of apertures into a pair of rectangular apertures with orthogonal in-plane orientation angles (a specific implementation of the *p2* plane group). The perturbation doubles the period along a real-space dimension and halves the first Brillouin zone, which folds a previously guided mode into the radiation continuum, forming a q-BIC mode. The Q-factor of the q-BIC is engineerable and scales inversely with perturbation strength δ, as $Q \propto \frac{1}{\delta^2}$.[29] In our devices, the perturbation strength is characterized by the aspect ratio of the rectangular aperture. We choose the geometrical parameters of our meta-units so that the q-BIC is a fundamental TE mode (**Fig. 2a**) in the near-infrared with substantial enhancement of the electric field in the germanium (**Fig. 2b**). We illuminate our nonlocal wavefront-shaping metasurfaces with circularly polarized light and measure transmitted circularly polarized light of the opposite handedness. Light only experiences polarization conversion and associated geometric phase when its spectrum overlaps with the narrowband q-BIC.[31] The geometric phase, $\phi$, is controlled by the in-plane orientation angle, α, of the rectangular apertures as $\phi \sim 4\alpha$ (**Fig. 2d**).[31] The simulated spectra for a representative meta-unit (**Fig. 2c**) demonstrate that adjusting the refractive index of the entire device by $\Delta n \approx 0.1$ is sufficient to shift the resonance by more than the linewidth of the resonant mode.

With this meta-unit library, we design a radial metalens with numerical aperture (NA) of ~0.1 and fabricate it following a fabrication procedure that requires no etching and only one pass of lithography (see **Materials and Methods**). The measured spectra of the fabricated device (**Figs. 2e** and **2f**) show a q-BIC mode with a Q-factor of ~80 that redshifts upon heating (**Fig. 2g**). We image handedness-converted circularly polarized light transmitted through the device at both the metasurface plane and the focal plane. At room temperature (23 °C), the device-plane (**Fig. 2h**) and focal-plane images (**Fig. 2i**) both show that the lens has a clear resonance centered at λ=1,540 nm: the zone pattern in the device-plane images is most evident near the resonant wavelength of the q-BIC due to the radiation of the q-BIC; the focal spot in the focal-plane images reaches peak intensity when the zone pattern is the strongest. At an elevated temperature of ~147 °C, the resonant wavelength redshifts by ~25 nm to λ=1,565 nm (**Figs. 2h and 2i**), which is consistent with the measured spectra in **Fig. 2g**. With incident light fixed at either λ=1,540 nm or 1,565 nm, lensing can be switched on and off by adjusting the temperature. This device has been heated up and cooled down repeatedly without noticeable degradation.

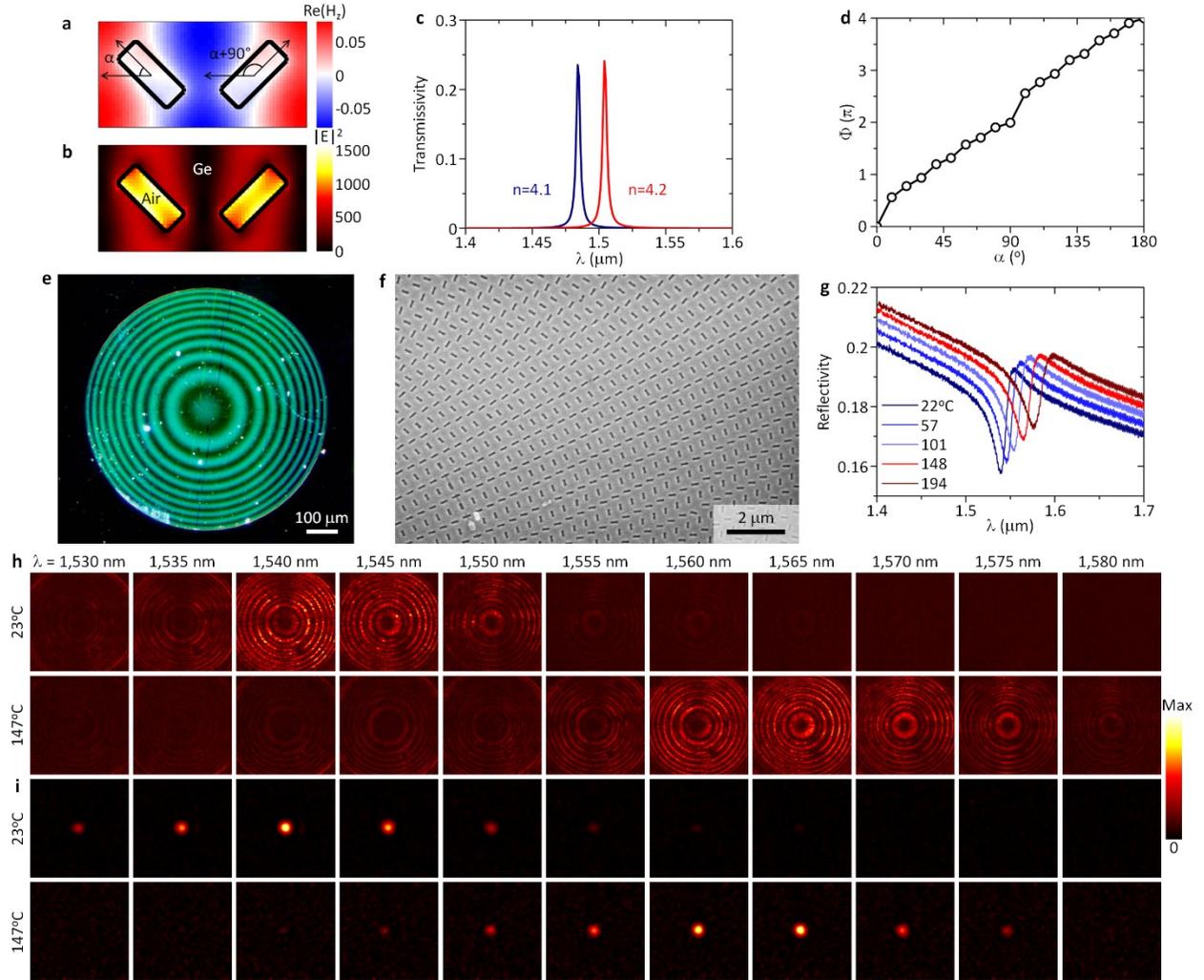

**Figure 2**. Design and experimental demonstration of a germanium metalens modulator. (a) Magnetic field profile of a meta-unit. (b) Electric field intensity profile of the meta-unit. (c) Simulated transmission spectra for circularly polarized light of converted handedness for the meta-unit with the refractive index of germanium being n=4.1 (blue) and n=4.2 (red). (d) Simulated geometric phase, $\phi$, as a function of the in-plane rotation angle, α, of the rectangular apertures at the resonant wavelength. (e) Dark field optical microscope image and (f) scanning electron micrograph of the fabricated metalens (Scale bar: 3 μm). (g) Measured temperature-dependent unpolarized reflection spectra of the metalens. (h-i) Measured, circularly polarized light of converted handedness at the device plane (h) and the focal plane (i) at two temperatures.

The dimensions of the device-plane images are 660 μm × 660 μm, and those of the focal-plane images are 330 μm × 330 μm. Device and meta-unit dimensions are summarized in **Table S2**.

We next demonstrate two devices in which the wavefront can be switched between that of two distinct cylindrical lenses by controlling the temperature. The devices are based on multi-perturbation nonlocal metasurfaces supporting two orthogonal q-BICs that each shape the wavefront at a different wavelength.[31] These metasurfaces are first illuminated with a narrowband light source tuned to the "red" (longer wavelength) resonant wavelength at room temperature, and then heated sufficiently so that the "blue" (shorter wavelength) resonant mode is redshifted to match with the wavelength of the incident light. In this way, the wavefront is switched between two distinct shapes encoded, respectively, in the geometric phase profiles of the two q-BIC modes.

The meta-unit geometry must be designed to support two orthogonal q-BICs, each capable of imparting a unique geometric phase profile, with the appropriate spectral spacing to allow for thermal switching of the two modes. We choose meta-units of rectangular apertures arranged into a particular configuration of a *p1* plane group so that two sets of orthogonal perturbations independently control the geometric phase for two different q-BICs. The two q-BICs of our choice are fundamental, anti-symmetric modes ($B_1$ modes) that are rotated 90° from each other (**Figs. 3b** and **3h**), and the spectral separation between the two modes is determined by the difference in the meta-unit lattice constant in the x and y directions. Specifically, increasing (decreasing) the lattice constant in only the x or y direction redshifts (blueshifts) the mode originating from dimerization perturbation along the same direction, with little impact to the other mode originating from perturbation in the orthogonal direction (**Fig. S2**). This direct control of the spectral mode spacing is advantageous for designing switchable metasurfaces with a given change in refractive index.

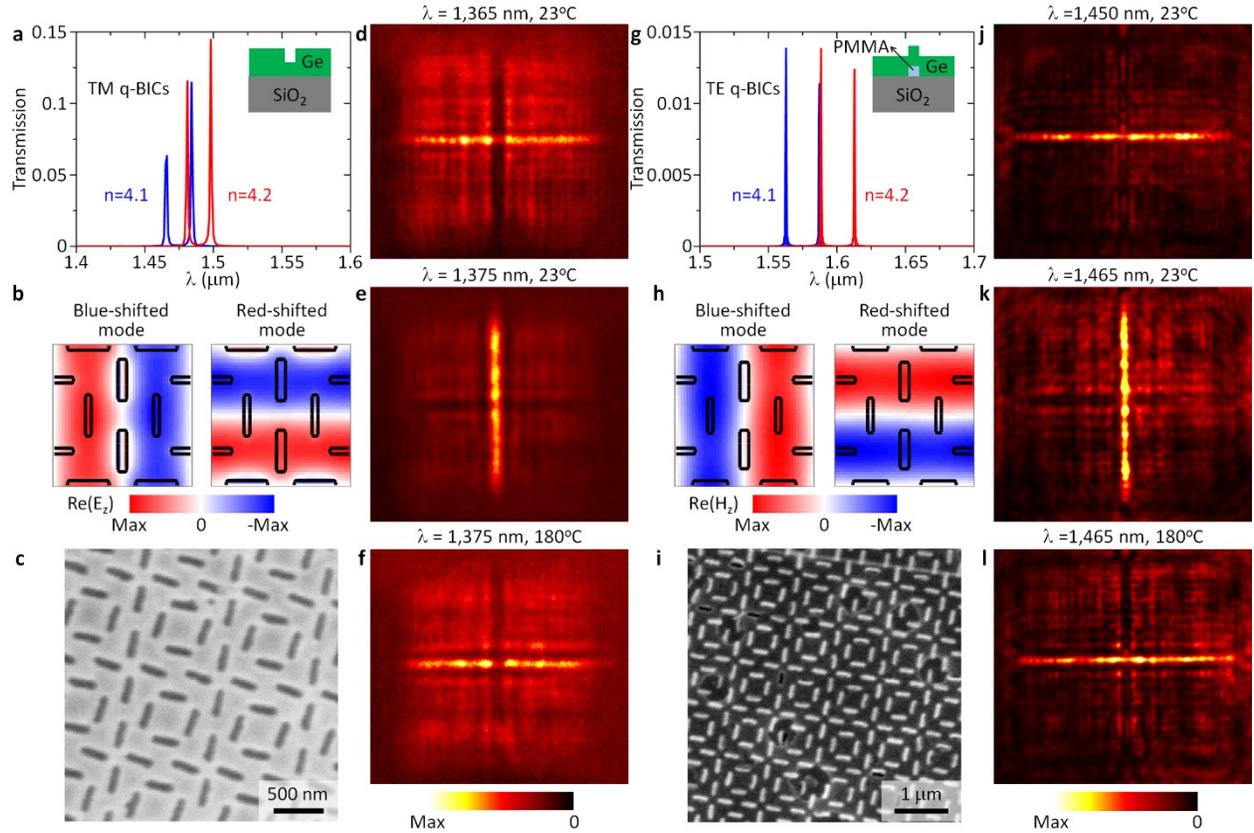

**Figure 3.** Design and experimental demonstration of two thermally switchable germanium metalenses. (a) Simulated transmission spectra of circularly polarized light of converted handedness for a meta-unit supporting two TM q-BIC modes. The blue and red spectra are calculated assuming that the refractive indices of germanium are 4.1 and 4.2, respectively. Inset to (a) depicts a vertical cross-section through the meta-unit. (b) Electric field ($E_z$) distribution over one meta-unit for the two TM q-BICs. Left: "blue" mode. Right: "red" mode. (c) Scanning electron micrograph of the TM-mode device. (d,e) Images of horizontal and vertical focal lines produced by the TM-mode device at λ=1,365 nm and 1,375 nm, respectively, at room temperature. (f) Image of the focal line produced by the TM-mode device at λ=1,375 nm and 180 °C. (g) Simulated transmission spectra of circularly polarized light of converted handedness for a meta-unit supporting two TE q-BIC modes. The blue and red spectra are calculated assuming that the refractive indices of germanium are 4.1 and 4.2, respectively. Inset to (g) depicts a vertical cross-section through the meta-unit. (h) Magnetic field ($H_z$) distribution over one meta-unit for the two TE q-BICs. Left: "blue" mode. Right: "red" mode. (i) Scanning electron micrograph of the TE-mode device. (j,k) Images of horizontal and vertical focal lines produced

by the TE-mode device at λ=1,450 nm and 1,465 nm, respectively, at room temperature. (l) Image of the focal line produced by the TE-mode device at λ=1,465 nm and 165 °C. Device and meta-unit dimensions are summarized in **Table S2**.

We demonstrate two switchable cylindrical metalenses: the first one supports two TM q-BICs (**Fig. 3b**) and consists of two sets of rectangular apertures part-way etched through a germanium thin film (inset to **Fig. 3a**); the second supports two TE q-BIC modes (**Fig. 3h**) and consists of two sets of rectangular PMMA rods covered with germanium (inset to **Fig. 3g**) . **Figure 3a** (**3g**) shows simulated transmission spectra of circularly polarized light with converted handedness showing the pair of TM (TE) modes and that they redshift to a similar degree when the refractive index of germanium increases. By independently controlling the orientations of the two sets of apertures or PMMA rods, we encode the "blue" mode (i.e., the one with shorter wavelength among the pair of q-BICs) with a spatial distribution of geometric phase representing a horizontal cylindrical lens (i.e., a lens that produces a horizontal focal line) and the "red" mode with a geometric phase profile representing a vertical cylindrical lens. Both lenses have NA of approximately 0.05. Scanning electron micrographs of the two metalenses are shown in **Figs. 3c** and **3i**, respectively. Imaging the far field of the TM-mode device at room temperature reveals a horizontal focal line at λ= 1,365 nm (**Fig. 3d**) and a vertical focal line at λ= 1,375 nm (**Fig. 3e**). We note that this device is deliberately tilted with respect to the incident beam to leverage the differential angular dispersion of the two q-BICs (i.e., different magnitude of resonant wavelength shift as a function of incident angle that is associated with the bandstructure of the nonlocal device) to decrease their spectral separation to only dλ~10 nm. For incident light at λ= 1,375 nm, the device switches from a vertical cylindrical lens to a horizontal cylindrical lens as it is heated to 180 °C (**Figs. 3e** and **3f** and Supporting Info Video). The far field of the TE-mode device shows a horizontal focal line at λ= 1,450 nm (**Fig. 3j**) and a vertical focal line at λ= 1,465 nm (**Fig. 3k**) for normal incidence. At λ= 1,465 nm, the metasurface changes from a vertical cylindrical lens to a horizontal one upon heating (**Fig. 3l**). The peculiar meta-unit structure in the second cylindrical metalens (**Figs. 3g-l**) results from a failed "liftoff" step during device fabrication. Nevertheless, the metalens functions properly in experiments, and simulations (**Fig. S1**) demonstrate that it operates similarly in terms of tunability to the initial design consisting of apertures in a germanium thin film. The wavefront-shaping q-BICs in our nonlocal metasurfaces arise from specific, deliberate perturbations to the in-plane symmetry of the PhC lattice. The device functions appropriately despite the fabrication error because the in-plane symmetry perturbations are preserved. This points to robustness of symmetry-protected q-BICs in nonlocal metasurfaces as a platform for switchable metasurfaces.

Thermo-optic switching of nonlocal metasurfaces made of conventional dielectric materials is a generalizable concept, and switchable devices can be readily realized in the visible wavelength range. **Figure 4** shows a design for a visible-spectrum, thermally-switchable beam steerer based on titanium dioxide ($TiO_2$), a popular material for passive local metasurfaces in the visible[36,37] with a thermo-optic coefficient of -0.5 ~ -2.1 ×10$^{-4}$ K$^{-1}$ [38,39]. In this multifunctional metasurface, we employ a *p1* meta-unit consisting of rectangular apertures in a thin film of $TiO_2$ and implementing two sets of perturbations, each supporting a q-BIC (TM mode with $B_1$ representation). By design, the two q-BICs are separated spectrally by only ~5 nm so that the device functionality may be switched by heating it to moderate temperatures. The in-plane rotation angles for each set of perturbations are arranged to form a constant phase gradient in the direction orthogonal to its dimerization direction (i.e., $\frac{d\phi}{dx} = \frac{4\pi}{1.624\ \mu m}$ for the "blue" mode and $\frac{d\phi}{dy} = \frac{4\pi}{1.6\ \mu m}$ for the "red" mode), as shown in **Fig. 4a**. In this way, the "blue" mode steers light along the x-direction and "red" mode the y-direction. As the devices is heated, the refractive index of the device decreases and the resonant wavelength of both q-BICs blueshifts (**Fig. 4b**). With the wavelength of incident light fixed at the resonant wavelength of the "blue" mode at room temperature, incident light is steered in the x-direction at room temperature (**Fig. 4c**) and in the y-direction at an elevated temperature (**Fig. 4d**).

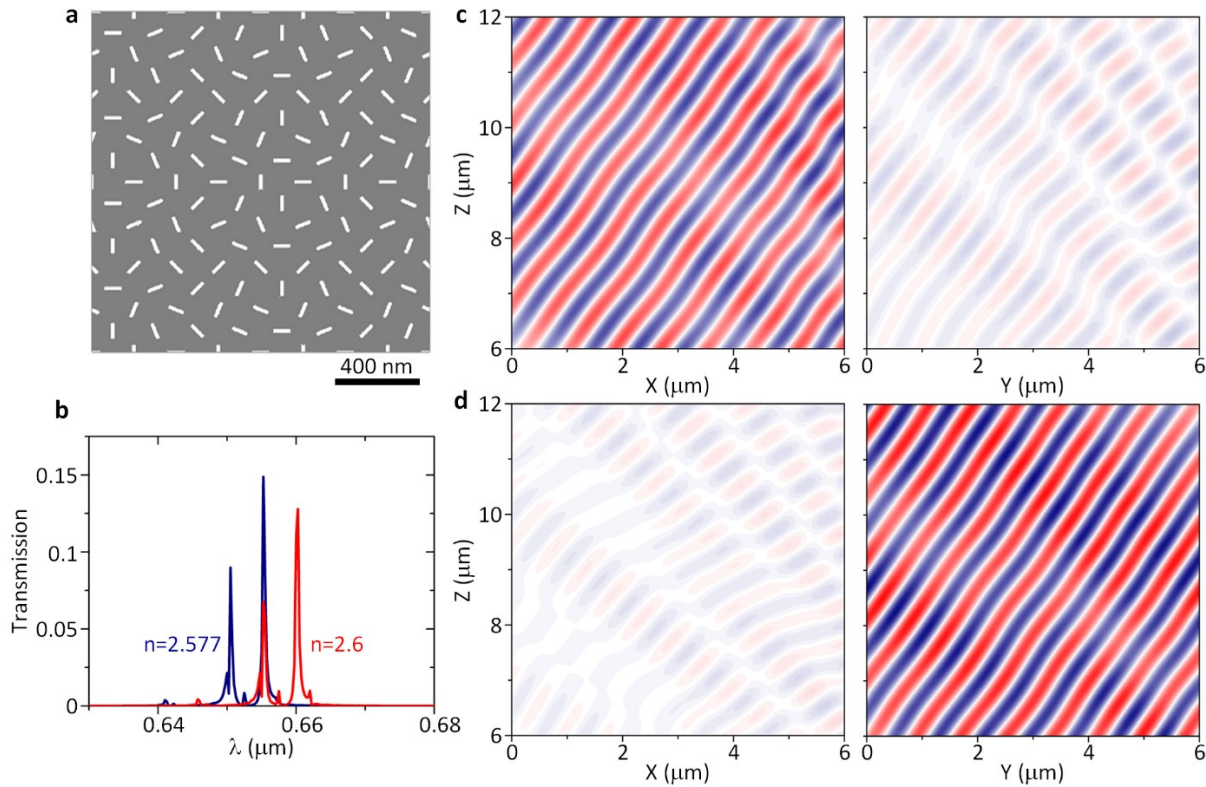

**Figure 4**. Design of a visible-spectrum, thermally-switchable metasurface based on titanium dioxide. (a) Schematic of a beam steering device consisting of apertures in a thin film of titanium dioxide. (b) Simulated transmission spectra of circularly polarized light of converted handedness for the device with $n_{TiO2}$=2.6 (black) and $n_{TiO2}$=2.577 (pink). (c-d) Simulated far field profiles at λ=655.3 nm of transmitted, circularly polarized light of converted handedness with $n_{TiO2}$=2.6 (c) and $n_{TiO2}$=2.577 (d). Meta-unit dimensions are summarized in **Table S2**.

Switchable devices in the visible are also realizable using silicon nitride despite its low thermo-optic coefficient (**Fig. S3**). Such devices require judicious engineering of the Q-factor and spectral mode separation (**Fig. S4**) but imply that a broad range of conventional materials can be employed as the active material in nonlocal metasurfaces. We have previously argued that nonlocal wavefront-shaping metasurfaces in the visible will prove useful for augmented reality (AR) headsets to focus contextual information to the viewer while simultaneously allowing an unimpeded view of the real world.[32] Implementing switchable nonlocal metasurfaces into an AR headset introduces additional flexibility and adaptivity.

In summary, we experimentally demonstrated thermo-optically reconfigurable nonlocal metasurfaces based on conventional CMOS-compatible dielectric materials and manufacturing processes. This device platform offers a means of introducing tunable metasurfaces into real-life applications by alleviating our dependence on novel tunable materials.[40] The platform is readily scalable in that highly multifunctional systems can be realized by cascading nonlocal metasurfaces, each supporting up to four distinct functionalities.[31,32] Therefore, tunable meta-optics systems capable of switching between many distinct wavefront shapes are feasible. The devices in this work provide a baseline of what is possible in active metasurfaces without resorting to exotic materials, unconventional fabrication procedures, or sophisticated device architectures. Admittedly, the thermo-optic effect used in these devices is a slow process and has high power consumption; we expect that tunable nonlocal metasurfaces made from electro-optic materials may be of use in the future for advanced applications requiring low power and fast switching speed.

**Materials and Methods**

**Fabrication** Fused silica substrates are washed and then spincoated with three sequential layers: PMMA A4 495 (4000 RPM for 45 seconds, followed by 5-minute bake at 180 °C), PMMA A2 950 (4000 RPM for 45 seconds, followed by 2-minute bake 180 °C), and an anti-charging layer disCharge H2O x2 (2000 RPM for 30 seconds). Devices are patterned with electron beam lithography (Elionix ELS-G100) at a current of 500 pA after proximity effect corrections are applied to the pattern file with the software program BEAMER. After exposure, the anti-charging layer is removed by rinsing in DI water, and the devices are developed in a chilled (6 °C) 3:1 isopropyl alcohol:deionized water solution for 2 minutes followed by 30 seconds of rinsing in deionized water. Germanium is deposited by electron beam evaporation (AJA International Orion-8E). The sample is then placed in N-Methyl-2-pyrrolidone (NMP) at 80 °C for at least four hours to remove the unpatterned PMMA and the germanium on top of it. The device in **Fig. 3a** is fabricated by depositing a layer of germanium on a fused silica substrate before beginning the above lithography process, deposition, and liftoff process in order to create a 'partially etched' metasurface.

**Optical Measurements** Samples are secured to a home-built resistive heating platform with the devices placed over a gap in the heater to allow for unimpeded transmission measurements. The temperature of the substrate is monitored by measuring the temperature of a dot of black paint on the substrate near the device with a thermal camera (FLIR T460). Spectrum measurements are made with a Fourier transform infrared (FTIR) spectrometer (Bruker Vertex 70v) and a mid-infrared microscope (Bruker Hyperion 2000). Far field measurements are made following the protocol described in previous work. [32] Briefly, incident light is prepared by coupling near-infrared light from a super continuum source (NKT Photonics SuperK Extreme) through a monochromator (Horiba iHR550) to a fiber collimator and then circularly polarized by a linear polarizer followed by a quarter-wave plate. Light emerging from the metasurface is collected by a 10× objective, one of its circularly polarized components is selected by another quarter-wave plate and linear polarizer, and then it is imaged by a near-infrared camera (NIRvana InGaAs camera, Princeton Instruments).

## Acknowledgements

This work was supported by the National Science Foundation (grant no. QII-TAQS-1936359 and no. ECCS-2004685) and the Air Force Office of Scientific Research (grant no. FA9550-14-1-0389 and no. FA9550-16-1-0322). Device fabrication was carried out at the Advanced Science Research Center NanoFabrication Facility at the Graduate Center of the City University of New York.